\documentstyle[11pt,moriond_hep,epsfig]{article}

\def\Journal#1#2#3#4{{#1} {\bf #2}, #3 (#4)}

\def\NPB{{\em Nucl. Phys.} B}
\def\PLB{{\em Phys. Lett.}  B}
\def\PRL{\em Phys. Rev. Lett.}
\def\PRD{{\em Phys. Rev.} D}

\def\be{\begin{equation}}
\def\ee{\end{equation}}
\def\bea{\begin{eqnarray}}
\def\eea{\end{eqnarray}}

\def\lsim{\mathrel{\raise.3ex\hbox{$<$\kern-.75em\lower1ex\hbox{$\sim$}}}}
\def\gsim{\mathrel{\raise.3ex\hbox{$>$\kern-.75em\lower1ex\hbox{$\sim$}}}}

\def\fbi{~{\rm fb}^{-1}}

\def\gev{\,{\rm GeV}}
\def\tev{\,{\rm TeV}}
\def\anti{\overline}
\def\hh{H}
\def\ha{A}
\def\tanb{\tan \beta}
\def\mhh{m_{\hh}}
\def\mha{m_{\ha}}

\begin{document}
\vspace*{4cm}
\title{TESTING CP VIOLATION IN THE HIGGS SECTOR AT MUON COLLIDERS$^\star$}

\author{B. GRZADKOWSKI}

\address{Institute of Theoretical Physics, Warsaw University, 
ul. Hoza 69, PL-00-681 Warsaw, Poland\\
E-mails: bohdang@fuw.edu.pl,pliszka@fuw.edu.pl}

\maketitle\abstracts{
We study the possibility of measuring the muon Yukawa couplings in 
s-channel Higgs boson production at a muon collider with transversely 
polarized beams. We investigate sensitivity to the relative size
of the CP-odd and CP-even muon Yukawa couplings. 
Provided the event rate observed justifies the operation of the
$\mu^+\mu^-$ Higgs boson factory, we find that 40\%
polarization for both beams is sufficient to resolve the CP nature
of a single resonance as well as disentangle it from two overlapping
CP conserving resonances.
}
\vspace*{\fill}
\footnoterule 
\vspace{2mm}
{\footnotesize $\;\;^\star$A report 
on work done in collaboration with J.F.~Gunion and J.~Pliszka.}

\newpage
\section{Introduction}

In spite of the fact that the Standard Model (SM) of electroweak interactions
has been tested with very high precision,
its scalar sector still evades experimental confirmation. In particular, it is an open question
if the proper theory should contain one or more physical Higgs bosons.
Since the CP nature of Higgs particles is a model dependent feature~\cite{hhg} its determination
would not only provide information concerning the mechanism of CP violation
but would also restrict possible extensions of the SM of electroweak interactions 
and therefore reveal the structure of fundamental interactions beyond the SM.
A muon collider with transversely polarized beams is the
only place where CP properties of a second generation fermion Yukawa coupling
can be probed. This is the subject of our
more complete analysis~\cite{hcpmupmum} which is summarized in 
this talk.
We follow the line of our previous works~\cite{gghgp}
where we have tried to unveil the CP-nature of Higgs bosons
in a model-independent way.

\section{Production of Higgs Bosons}

The attractive possibility of  s-channel Higgs boson production
at a muon collider has been discussed before~\cite{bbgh}
together with the possibility of the measurement of CP violation in the 
muon Yukawa couplings~\cite{bbgh,soni,hcpmupmum}.
The latter is based on the fact that 
in any muon collider design~\cite{pstareport,euroreport}
there is a natural beam polarization of the order of 20\%~\cite{rajat}
that allows for the rare possibility of direct Higgs boson production 
with known polarization of the initial state particles.

The cross section for the Higgs boson resonance production, $\mu^+\mu^-\!\rightarrow\!R $,
depends on the transverse $P_T^\pm$ and longitudinal $P_L^\pm$
beam polarizations and the $\bar\mu r \exp( i \delta\gamma_5)\mu$ 
muon Yukawa coupling in the following way:
\begin{equation}
\sigma_S(\zeta)=
\sigma_S^0 \left[1+P_L^+P_L^-+P_T^+P_T^-\cos(2\delta+\zeta)\right]
\label{sigform}
\end{equation}
where $\zeta$ is the angle in the transverse plane between 
the beam polarizations  and  $\sigma_S^0$ is the unpolarized cross section.
We stress that only the transverse polarization term is sensitive to 
$\delta$ of the muon Yukawa coupling. Since it is proportional
to the product of the transverse polarizations,
it is essential to have large $P_T^+$ and $P_T^-$, as obtained
by applying stronger cuts while selecting
muons from the decaying pions initially produced 
(which, however, causes a reduction of luminosity).
To compensate, a more 
intensive proton source or the ability to repack muon bunches will be needed.
Another  speculative option, would be high, up to 50\%,
polarization obtained by a phase-rotation technique~\cite{kaplan}
which would lead to less luminosity reduction.

While varying $\zeta$,  one can observe a maximum
at $\zeta=-2\delta$ and a minimum at $\zeta=\pi-2\delta$.
Thus, studying $\zeta$ dependence is essential for resolving the $\delta$
value. A muon collider offers the unique possibility of a setup which
in a natural way provides a scan over different $\zeta$
values. We will not discuss this option here. 
Our results will correspond to a configuration with four
fixed $\zeta$ values: 0,90,180 and 270 degrees.
Even though this cannot be accomplished experimentally,
due to the spin precession in the accelerator ring, 
it can be well approximated by a simple but realistic setup~\cite{hcpmupmum}
that yields the same results as the fixed $\zeta$ analysis
at the expense of 50\% luminosity increase.

\begin{figure}
\psfig{figure=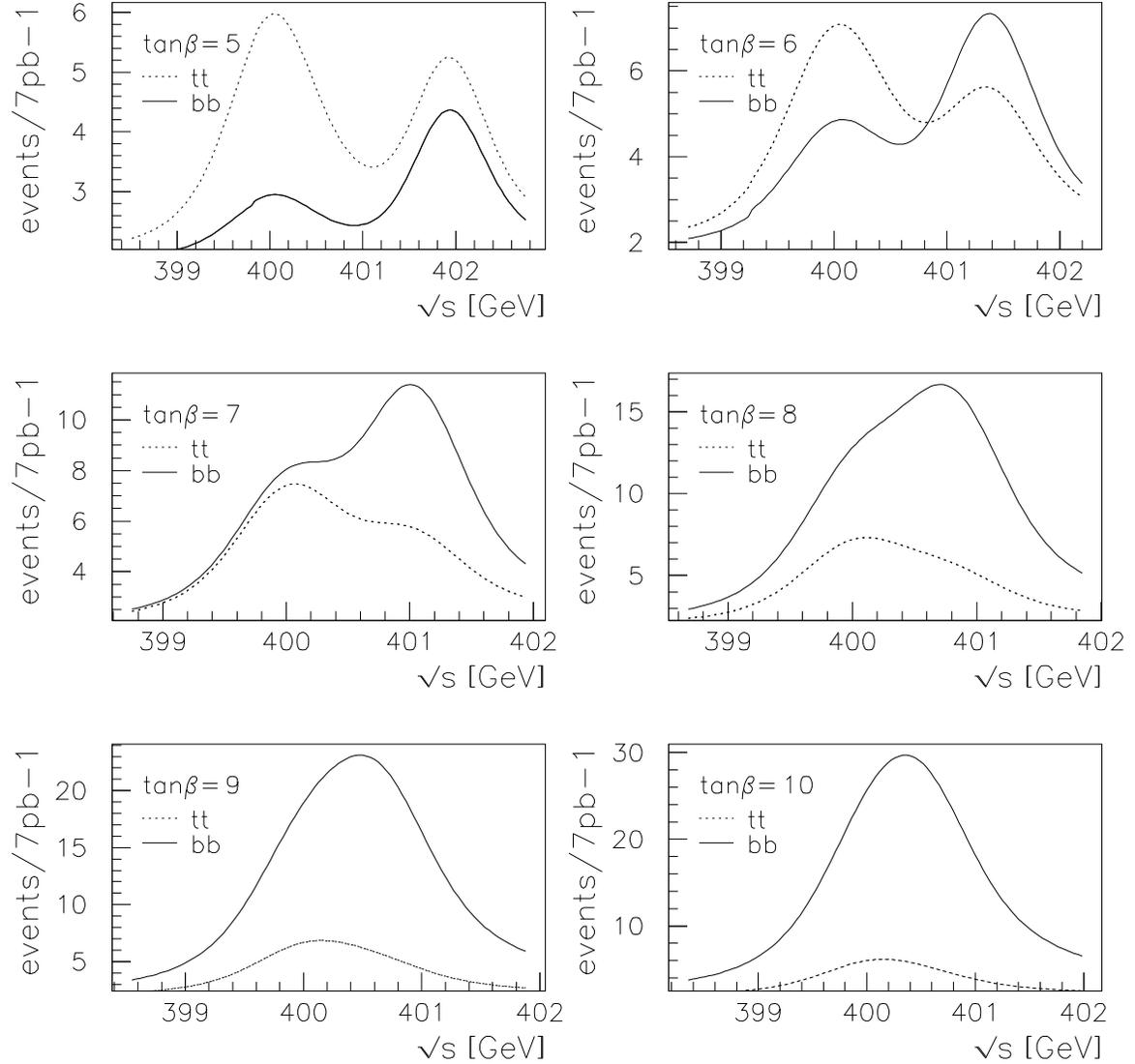, height=17cm}
\caption{We plot $b\bar{b}$ (solid) and $t\bar{t}$ (dashed) event rates
for total integrated luminosity of $L=7~{\rm pb}^{-1}$ coming from
$\mu^+ \mu^- \rightarrow H + A$ as a function of $\protect\sqrt{s}$, assuming 
$m_A=400\;\textrm{GeV}$.
Each window is for the specific $\tan\beta$ value noted. These event
rates are to be multiplied by a factor of 1000 for the expected
yearly integrated luminosity of $7~{\rm fb}^{-1}$ for the beam energy spread $R=0.1\%$. 
We employ squark masses
of $1\;\textrm{TeV}$ and no squark mixing. Supersymmetric decay channels are
assumed to be closed.}  
\label{degeneracy}
\end{figure}

\begin{table}[h]
\caption{Assuming the SM with $\delta=0$, we give 1 and 3 $\sigma$  
limits (in radians) on $\delta$ for the $b\bar{b}$ 
final state for various luminosity and polarization configurations.
Beam energy spread 
0.003\% and $b\bar{b}$ tagging efficiency  $54\%$ have  been assumed.}
\vspace{.4cm}
\begin{center}
\begin{tabular}{|c|c|c|c|c|c|}
\cline{3-6}
\multicolumn{2}{c|}{} & 
\multicolumn{2}{|c|}{$m_R$=110 GeV} & 
\multicolumn{2}{|c|}{$m_R$=130 GeV} \\
\hline
P[\%] & L[pb$^{-1}$] & 1$\sigma$& 3$\sigma$& 1$\sigma$& 3$\sigma$\\
\hline
20 & 150 & 0.94 & -- &  -- & -- \\
39 & 75  & 0.30 & 1.14 & 0.41 & --\\
48 & 75  & 0.20 & 0.64 & 0.27 & 0.93 \\
45 & 150 & 0.15 & 0.50 & 0.21 & 0.69\\
\hline
\end{tabular}
\end{center}
\label{tab_results}
\end{table}
In order to illustrate the ability to reject different Higgs boson
CP scenarios we can assume that the measured data is mimicked by 
the SM Higgs boson. 
For given luminosity $L$ and {\it total} polarization $P$ for each
of the beams, we can place
1 and 3 $\sigma$ limits on the $\delta$ value for the observed resonance,
assuming the $\delta=0$ SM is input.
The limits for Higgs boson masses  of 110 and 130 GeV are presented
in table~\ref{tab_results}. For the expected yearly luminosity of
$L=150~{\rm pb}^{-1}$, even several years of running
at the natural 20\% polarization would be insufficient for useful limits. 
However, 1$\sigma$ limits for the $P=39\%$ option (with reduced $L$) 
do give a rough indication of the CP nature of the resonance.
3$\sigma$ limits 
in 110-130 GeV mass range require either $>40\%$ polarization 
or $<50\%$ luminosity loss.
(The requirements are less stringent for a 110 GeV Higgs boson.)
{\bf We stress that there is no other way the measurement 
of the muon Yukawa $\delta$ can be done
and that operation in the transverse polarization mode should not
interfere with most of the other studies.}
For a heavier resonance, operation of a muon collider as an s-channel
Higgs boson factory is justified only if the branching ratio
$BR(R\rightarrow \mu^+\mu^-)$ is enhanced. Then, the analysis 
sketched above applies as well.
Such enhancement arises in the Minimal Supersymmetric 
Standard Model (MSSM) at large $\tan\beta$.
If the pseudoscalar mass is large ($m_A >300\;\textrm{GeV}$), the
$H$ and $A$ masses will be similar. 
The increasing degeneracy with increasing $\tan\beta$ is illustrated
for $m_A=400\;\textrm{GeV}$ in Fig.~\ref{degeneracy},
assuming squark masses of $1\;\textrm{TeV}$ and no squark mixing and a 
beam energy spread of $R=0.1\%$. 
Since the total widths of the $H$ and $A$
are substantial ($>1\;\textrm{GeV}$) for the $m_A$
and $\tan\beta$ values being considered, it is not
guaranteed that we will be able to separate the peaks.
The figure shows that we are able to observe two separate peaks (the $A$
peak being at lower mass than the $H$ peak) for moderate
$\tan\beta\lsim 6$.  But, for higher $\tan\beta$ values
the peaks begin to merge; for $\tan\beta\gsim 8$, $|m_H-m_A|<11\;\textrm{GeV}$ and one
sees only a single merged peak.  The picture changes if squark mixing
is substantial; for instance, for $m_A=300\;\textrm{GeV}$, squark
masses of $1\;\textrm{TeV}$ and large squark mixing ($A_t=A_b=3\;\textrm{TeV}$), 
the $H$ and $A$ peaks actually cross at $\tan\beta\sim 5$. 
It would be crucial to distinguish such a case from a single
CP-violating Higgs boson which may appear e.g. in the MSSM~\cite{cp-phases}. 
Table \ref{patterns} illustrates
the very distinct event number pattern as a function of $\zeta$
that would yield the needed discrimination.
The event rate for any single, CP conserving or CP violating Higgs boson,
has a minimum and maximum as a function of $\zeta$. In contrast,
overlapping CP-even and CP-odd resonances
result in a pattern independent of $\zeta$.
\begin{table}[h]
\vspace{-4mm}
\caption{Event number pattern for different Higgs models as
a function of $\zeta$, assuming $P_L^\pm=0$ and $P_T^\pm=P$; see
Eq.~(\ref{sigform}).}
\vspace{0.4cm}
\begin{center}
\begin{tabular}{|c|c|c|c|c|}
\hline
\raisebox{0pt}[12pt][7pt]{$(\delta)$} & $\zeta=0$ & 
$\zeta=\pi/2$  & $\zeta=\pi$ & $\zeta=3\pi/2$ \\
\hline
\raisebox{0pt}[10pt][5pt]{$(0)$} & $1+P^2$ & 1 & $1-P^2$ & 1 \\
\raisebox{0pt}[5pt][5pt]{$(\pi/4)$} & 1 & $1-P^2$ & 1 & $1+P^2$ \\
\raisebox{0pt}[5pt][5pt]{$(\pi/2)$} & $1-P^2$ & 1 & $1+P^2$ & 1 \\
\raisebox{0pt}[5pt][5pt]{$(0)+(\pi/2)$ }& 1 & 1 & 1 & 1 \\
\hline
\end{tabular}
\end{center}
\label{patterns}
\end{table}

\begin{figure}
\psfig{figure=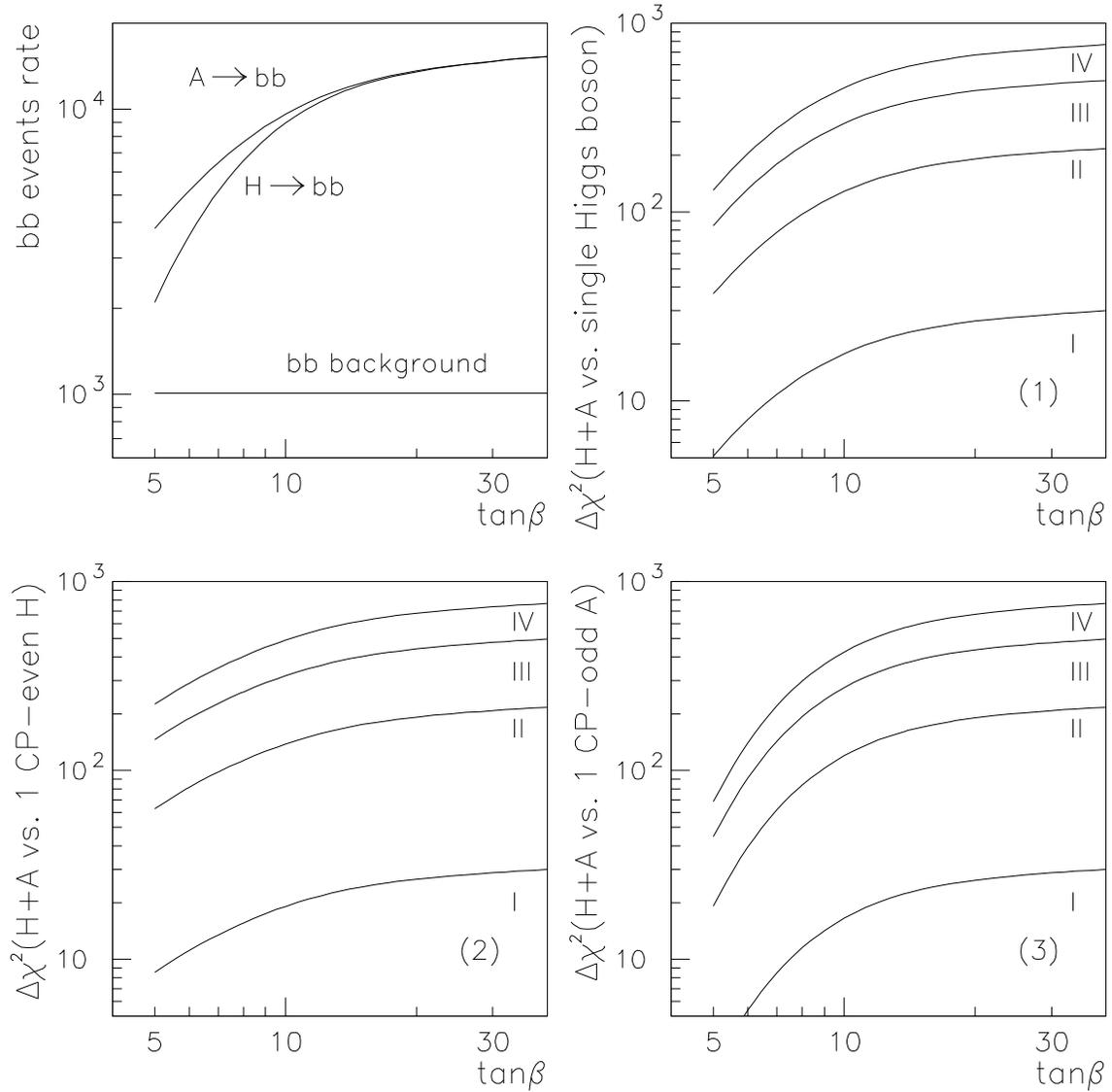, height=17cm}
\caption{
In the upper left  window, we plot the $b\anti b$ event rates
in the MSSM for the $\hh$ and $\ha$ (the $\ha$ rate is the larger
of the two) as a function of $\tanb$ for $\mha=300\gev$,
assuming squark masses of $1\tev$, no squark mixing and integrated
luminosity of $L=2\fbi$.
Also shown is the (relatively small) background rate.
In the remaining windows we plot $\Delta\chi^2$,
after including precession and increasing $L$ to $L=3\fbi$, 
as a function of $\tanb$ for three different cases
in which we forcibly lower $\mhh$ to $300\gev$ (for exact degeneracy).
(1) We adjust the $\hh$ and $\ha$ event rates so that each is exactly equal
to the average of the $\ha$ and $\hh$ rates
as predicted by the MSSM, and compute $\Delta\chi^2$ for $\hh+\ha$ vs.  
a single Higgs resonance (of any type) with the same total 
event rate, employing only the $b\anti b$ channel. 
(2) We use the actual $\hh$ event rate and compute
$\Delta\chi^2$ for $\hh+\ha$ vs. a single CP-even resonance
yielding the same $b\anti b$ event rate.
(3) As in (2), but vs. a single CP-odd resonance.
In cases (1)-(3), we give results as a function of $\tanb$
for the four polarization--luminosity
situations (I)-(IV) (as labelled on the curves) described in the text.}
\label{hh300}
\end{figure}

In Fig.~\ref{hh300} we plot for $m_A=300\gev$
$\Delta\chi^2$~\footnote{To test a model A against B we introduce 
$\Delta\chi^2=\Sigma_i \frac{(N^A_i-N^B_i)^2}{{N^B_i}}$, where 
$N^{A/B}_i$ denotes the number of events in the $i$th bin calculated 
within the model $A/B$; see Grzadkowski, Gunion and Pliszka~\cite{hcpmupmum} for details.
The background generated by $\gamma^\star$ and $Z$ exchange is 
taken into account.}
obtained for the four polarization-luminosity options (I)-(IV): 
(I)   $P=0.2$,  $L=3.0\fbi$,
(II)  $P=0.39$, $L=1.5\fbi$,
(III) $P=0.48$, $L=1.5\fbi$,
(IV)  $P=0.45$, $L=3.0\fbi$.
We emphasize that (for $R=0.1\%$) options (I) and (II) do not require over-design of the proton
source. The $\Delta\chi^2$ plots show that good discrimination is obtained
even for option (I) once $\tan\beta>10$. Option (II)  would be
needed for good discrimination if $\tan\beta\sim 5$.

We have found that for simple MSSM test cases with $m_A= 300-400$ GeV
and $\tan\beta> 8$ (for which we cannot see separate resonance peaks) 
even natural 20\% polarization
will allow us to distinguish two overlapping resonances from any single one
at more than the 3$\sigma$ level.
Higher polarization will allow for a precise measurement
of the relative contribution from the CP-even and the CP-odd component.

\section{Summary and Conclusions}

We have presented results of 
a realistic study of measuring the
CP properties of the muon Yukawa couplings in 
Higgs boson production at a muon collider with transversely polarized beams.
We have found that transverse polarization is essential for determining
the CP nature of the 
muon Yukawa couplings. In particular, a collider with 
$P\!\approx\!40\%$ and at least 50\% of the 
original luminosity retained 
will ensure that the CP nature of the produced scalar resonance will be revealed. 

\section*{Acknowledgments}
The author is grateful to the organizers of the XXXVth Rencontres de Moriond, 
on ``Electroweak Interactions and Unified Theories'' for creating a very warm 
and inspiring atmosphere during the meeting. He thanks J.F.~Gunion and J.~Pliszka 
for a collaboration upon which the presented talk was based and 
S.~Geer, R.~Raja and R.~Rossmanith for helpful conversations
on experimental issues.
This work was supported in part by the U.S. Department of Energy,
the U.C. Davis Institute for High Energy Physics, the State Committee for
Scientific Research (Poland) grant No. 2~P03B~014~14 and by Maria
Sklodowska-Curie Joint Fund II 
(Poland-USA) grant No. MEN/NSF-96-252.

\end{document}